\documentclass[]{gTHM2e}
\usepackage{amsmath, amssymb, mathrsfs,ulem}
\usepackage{color}
\usepackage{graphicx}
\newcommand{\RR}{\mathbb{R}}
\newcommand{\ud}{\mathrm{d}}
\begin{document}

\markboth{S. E. Wang et al.}{A mathematical model quantifies proliferation and motility effects of TGF--$\beta$ on cancer cells}
\title{A mathematical model quantifies proliferation and motility effects of TGF--$\beta$ on cancer cells}

\author{Shizhen Emily Wang$^\textrm{a}{}^\ast$ \thanks{$^\ast$ These authors contributed equally to this study.}  
Peter Hinow$^\textrm{b}{}^\ast$ 
Nicole Bryce$^\textrm{a}{}^\ast$ 
Alissa M. Weaver$^\textrm{a}$ 
Lourdes Estrada$^\textrm{a}$ 
Carlos L. Arteaga$^\textrm{a}$
and Glenn F.~Webb$^\textrm{c}$ $^\dagger$ \thanks{$^\dagger$ Corresponding author. Phone: +1 615 322 6661, Fax: +1 615 343 0215, E-mail: \texttt{glenn.f.webb@vanderbilt.edu}}\\\vspace{6pt} 
$^\textrm{a}$\textit{Department of Cancer Biology, Vanderbilt University, Nashville, TN 37232, USA} 
$^\textrm{b}$\textit{Institute for Mathematics and its Applications, University of Minnesota, 114 Lind Hall, Minneapolis, MN 55455, USA}\\$^\textrm{c}$\textit{Department of Mathematics, Vanderbilt University, Nashville, TN 37240, USA} 
}
\maketitle

\begin{abstract} Transforming growth factor (TGF) $\beta$ is known to have properties of both a tumour suppressor and a tumour promoter. While it inhibits cell proliferation, it also increases cell motility and decreases cell-cell adhesion. Coupling mathematical modeling and experiments, we investigate the growth and motility of oncogene-expressing human mammary epithelial cells under exposure to TGF--$\beta$. We use a version of the well-known Fisher--Kolmogorov equation, and prescribe a procedure for its parametrisation. We quantify the simultaneous effects of TGF--$\beta$ to increase the tendency of individual cells and cell clusters to move randomly and to decrease overall population growth. We demonstrate that in experiments with TGF--$\beta$ treated cells \textit {in vitro}, TGF--$\beta$  increases cell motility by a factor of 2 and decreases cell proliferation by a factor of 1/2 in comparison with untreated cells. \\

\noindent \textbf{Keywords: } Transforming growth factor (TGF) $\beta$, cell growth and motility,  mathematical model,  Fisher--Kolmogorov equation \\

\noindent\textbf{AMS Subject Classification:} 92B05, 92C37
\end{abstract}

\section{Introduction to the Biology of TGF--$\beta$}
In normal organisms, the growth of cells is under tight
regulation by growth factors and is highly dependent on the developmental
stage during the lifespan of the organism. Disruption of this regulation
is the most frequent cause of cancer diseases. Unlike
normal differentiated cells, cancer cells are usually
hyperproliferative as a result of the abnormal activation of
multiple growth--stimulating intracellular signalling pathways and
loss of tumour suppressors. In cancer cells, these pathways do not
respond to normal regulatory signals but are manipulated by one or
more oncogenic signals, often encoded by oncogenes. Expression of
oncogenes alters signalling pathways that under normal conditions
maintain cell growth homeostasis. Thus, altering these pathways may
favour increased cell and cancer growth. A good example is the
transforming growth factor (TGF) $\beta$ family, which is known to
be able to act as both a tumour suppressor and tumour promoting factor.

The TGF--$\beta$ family consists of multitasking cytokines that
play important roles in cell proliferation, cell motility,
apoptosis, lineage determination, extracellular matrix production,
and modulation of immune function \cite{1}. These ligands bind to a
heteromeric complex of transmembrane serine/threonine kinases, the
type I and type II receptors (T$\beta$RI and T$\beta$RII). The
receptors are activated upon ligand binding, leading to the
subsequent phosphorylation and activation of a family of
transcription factors called Smads, which regulate transcription of
a subset of genes \cite{3}. In addition to Smads, other signalling
pathways have been implicated in TGF--$\beta$ actions in recent
studies. These include the extracellular signal--regulated kinase
(ERK, MAPK), c--Jun NH2--terminal kinase (JNK), p38MAPK,
phosphatidylinositol--3 kinase (PI3K), and Rho GTPases (reviewed in
\cite{4,5,6,7}). The critical role of these non--Smad pathways on
mediating the cellular effects of TGF--$\beta$ remains to be fully
characterised.

TGF--$\beta$ was originally reported to induce transformation of
mouse fibroblasts \cite{8}. Subsequent studies indicated that
TGF--$\beta$  is a potent inhibitor of cell proliferation and a
tumour suppressor \cite{9,10}. Consistent with its tumour suppressor
role, many cancers lose or attenuate TGF--$\beta$--mediated
anti--mitogenic action by mutational inactivation of TGF--$\beta$
receptors or their signal transducer Smads \cite{11,12,13,14,15,16}.
There is increasing evidence to show that excess production
and/or activation of TGF--$\beta$ in tumours can accelerate cancer
progression through enhancement of tumour cell motility and survival,
increase in tumour angiogenesis, extracellular matrix production and
peritumoural proteases, and the inhibition of immune surveillance
mechanisms in the cancer host (reviewed in \cite{4,5,17}). Cancer progression and metastasis consist of a series of sequential events. After initial cell transformation, often mediated by the function of oncogenes, tumour cells growing at the primary site will invade the surrounding stroma and migrate towards blood vessels. Through various mechanisms such as epithelial--mesenchymal transition (EMT), tumour cells will enter the blood vessels and travel to other parts of the body through the circulatory system. Some of the cells will then arrest at distant sites where they may proliferate and invade into the adjacent organs. Cell motility is therefore a critical element during the spread of tumour cells from their initial sites of residence. In this study, we focus on the tumour--promoting effect of TGF--$\beta$ through inducing cell motility.

The receptor tyrosine kinase HER2 (ErbB2, Neu) belongs to the family
of epidermal growth factor receptor (EGFR). Gene amplification or
overexpression of HER2 is observed in about 25\% of breast cancers.
TGF--$\beta$ has been shown to synergize with the oncogene ErbB2 in
cancer progression. Overexpression of active TGF--$\beta$ 1 or
active mutants T$\beta$RI (Alk5) in the mammary gland of bigenic
mice also expressing mouse mammary tumour virus (MMTV)/Neu (ErbB2)
accelerates metastases from Neu--induced mammary cancers
\cite{18,19,20,21}. Exogenous as well as transduced TGF--$\beta$
confer motility and invasiveness to MCF10A non--transformed human
mammary epithelial cells (HMEC) stably expressing transfected HER2
\cite{25,26}. Expression of the oncogene HER2 in these cells does
not affect the function of TGF--$\beta$  on inhibiting cell
proliferation \cite{26}. It is likely that in many cancers,
TGF--$\beta$  may still attenuate proliferation while inducing
cellular events associated with metastatic dissemination, such as
cell motility. In this paper we report experiments with MCF10A/HER2 cells to study
and to separate the effects of TGF--$\beta$  on cell proliferation
and motility. Due to the complexity of TGF--$\beta$ signalling that
simultaneously affects several biological parameters, it is
important to computationally simulate the behaviour of cells under
TGF--$\beta$ exposure. Our model, which can also be adopted to
simulate other growth--regulating signals, will provide a unique
insight into the TGF--$\beta$ function in both normal and cancer
cells, and further understanding of targeted therapeutic strategies
that aim at interfering with TGF--$\beta$ signalling.

Mathematical modeling of chemotherapy with Tamoxifen has been carried out in \cite{Turner}, with special consideration of TGF--$\beta$. There, the authors use a discrete cellular automaton (CA) model to investigate the effects of TGF--$\beta$ on tumour morphology and invasiveness. It is possible to assign properties to individual cells in a cellular automaton model (such as different levels of malignancy). However, they can be computationally costly, in particular in situations where large cell populations are involved. Since we are interested chiefly in the effects of TGF--$\beta$ on a large number of cells growing in a plate, rather than individual cell behaviour, we work with a continuous partial differential equation model. Future research problems may require the construction of hybrid models such as in \cite{Anderson} where discrete cells are coupled to continuous fields that in turn are solutions of  partial differential
equations.

\section{Introduction to the Mathematical Model}
The Fisher--Kolmogorov equation (or KPP equation) has been studied and used widely in mathematical biology. It originated in the 1930s in works of R.~A.~Fisher and A.~N.~Kolmogorov \cite{Fisher,Kolmogorov} where it was proposed as a model for the spread of an advantageous gene in a population.  Mathematically, it falls into the large class of reaction--diffusion equations, in which one or more diffusible ``species'' enter into a scheme of reactions. The species can be chemical substances or  biological populations. Reaction--diffusion equations in general and the Fisher--Kolmogorov equation in particular have been applied successfully by many authors to model the behaviour of cells in tissues, see for example \cite{Ayati,Anderson,Chaplain98,Enderling,Chaplain_Matzavinos,Drasdo,Lefever,Prigogine}. The Fisher--Kolmogorov equation is usually used to show traveling wave behaviour, but we use it here to model the  spatial growth of a cell culture from initial seeding to confluence.

Let $u(x,y,t)$ denote the spatial density of tumour cells at time $t$ in a dish corresponding to a spatial region $\Omega \subset \RR^2$. The time evolution of this density is described by the
partial differential equation
\begin{equation}\label{KPP}
\frac{\partial}{\partial t}u(x,y,t)= D\Delta u(x,y,t) + \alpha u(x,y,t)(1-\beta
u(x,y,t)).
\end{equation}
Here $\alpha$ incorporates the intrinsic proliferation rate (unit \textit{time}$^{-1}$), from initial seeding to confluence, in a single value independent of $x,y,t$. Also, $\beta$ denotes the area occupied by an average single cell. The density satisfies, $u(x,y,t)\leq\beta^{-1}$ for all $x,y,t$, and the total number of cells in the region $\Omega\subset\RR^2$ at time $t$
is given by
\begin{equation*}
N(t) = \int\int_\Omega u(x,y,t)\,\ud x\,\ud y.
\end{equation*}
The spatial region $\Omega$ may be the entire region occupied by the population or a subfield thereof. Equation \eqref{KPP} is accompanied by an initial condition and  appropriate boundary conditions. If a subfield $\Omega_1$ of $\Omega$ is completely populated, then
$u(x,y,t)=\beta^{-1}$ for all $x,y  \in \Omega_1$, otherwise $u(x,y,t)\leq\beta^{-1}$ and
\begin{equation*}
\frac{\beta}{|\Omega_1|} \int\int_{\Omega_1} u(x,y,t)\,\ud x\,\ud y
\end{equation*}
is the fraction of possible cells in $\Omega_1$ at time $t$. Here and in the sequel $|\cdot|$ denotes the area of the domain in question. We denote the total mass of the density by
\begin{equation}\label{mass}
U(t) = \beta \int\int_\Omega u(x,y,t)\,\ud x\,\ud y.
\end{equation}
Much of the remainder of this paper will be concerned with the
problem of finding parameter values to a modified version of equation
\eqref{KPP} from experimental observations and with the
interpretation of the solution to the equation.

The diffusion constant $D$ in equation \eqref{KPP} has to account for three simultaneous processes contributing to the spatial movement of cells. First, dividing cells occupy increasing space, therefore a certain part of $D$, say $D_p$, has to
account for the spatial expansion of proliferating cells. The authors of \cite{Chaplain_Matzavinos,Drasdo,Lefever,Prigogine} have proposed relations of the type
\begin{equation*}
D_p = k_1 \frac{\ell^2}{T_d}.
\end{equation*}
Here $\ell$ is related to the typical length or diameter of a cell and  $T_d$ is the doubling
time. The dimensionless factor $k_1$ in \cite{Drasdo} accounts for the expansion of the viable rim of dividing cells. We take $\ell$ to be the average increase in cell diameter between mitotic events. We assume the area of an average cell increases from approximately 50$\, \mu m^2$ to 100$\, \mu m^2$, so we take $\ell=3.3 \, \mu m$. We assume the average cycle time is $T_d=16\,h$.
We take $D_p = 0.1 \, \mu m^2 h^{-1}$ (order of magnitude).  Second, individual cells plated on a Petri dish do not remain fixed in their position but undergo a random walk. It is natural to assume that the random walk is not biased and follows the laws of Brownian motion. Third, cells in a cluster may break lose from that cluster as TGF--$\beta$ is known to decrease cell--cell adhesion \cite{28}. Thus, a part of $D$, say $D_m$, must account for this second and third contribution to cell movement.

We propose here a modification of the standard Fisher--Kolmogorov equation \eqref{KPP}. Our reasoning is that as the cells become more densely packed, their random motility should decrease. Thus, with $U$ from equation \eqref{mass} we propose
\begin{equation}\label{KPP2}
\begin{aligned}
\frac{\partial}{\partial t}u(x,y,t)&= \nabla\cdot(D(U)\nabla u(x,y,t)) +
\alpha u(x,y,t)(1-\beta u(x,y,t)), \\
D(U) &= D_m\left(1-\frac{U}{|\Omega|}\right)+ D_p.
\end{aligned}
\end{equation}
The constant $D_m$,  which captures the random motility of individual cells and clusters and the tendency of clusters to break apart, depends on TGF--$\beta$ concentration. We remark here that $D$ could also be made dependent on the local density $u$. In the case of interest to us in the present paper the initial seeding of the cells results in a uniform dispersion of the population that justifies the approximation made in equation \eqref{KPP2}. Since $D_p>0$, $D(U)>0$ and the modified equation \eqref{KPP2} has the important property of mathematical well--posedness. In practical applications, $D_p$ is often considerably smaller than $D_m$ (see the Results section and table \ref{Tab2} for our numerical parameter values). 

Different numerical values for the constant $D_m$ have been proposed
in the literature. Bray reports in his 1992 book
\cite[Table 1--1]{Bray} $D_m = 5\cdot10^{-10}\,cm^2\,s^{-1}=180\,\mu
m^2 h^{-1}$. Chaplain and Matzavinos \cite{Chaplain_Matzavinos}
arrive at a similar value $D_m = 7\cdot
10^{-5}\,cm^2\,day^{-1}=300\,\mu m^2 h^{-1}$. This value is obtained
in \cite{Chaplain_Matzavinos} from the Einstein--Stokes equation \cite{Berg}
\begin{equation*}
D=\frac{k_BT}{3\pi d\eta},
\end{equation*}
where $k_B$ is Boltzmann's constant, $T$ the temperature, $d$ the diameter of the cells, and $\eta$
the dynamic viscosity of the surrounding medium. It can be
excluded, however, that cells  \textit{in vitro}, such as in our experiments, move in an
environment that has the viscosity of water, $\eta_{H_2O} = 10^{-3}\, Pa\, s$. Thermally driven motion and Stokes friction are not relevant in the context of moving cells.

\section{Materials and Methods}

MCF10A/HER2 cells were generated and grown as described \cite{26}.
All live cell imaging experiments were performed in triplicate.
For growth assay, equal number of cells ($1.5 \cdot 10^4$/well)
suspended in full medium were seeded on 6--well plates. Cells were
allowed to grow in the absence or presence of TGF--$\beta$ at
different concentrations ($0.5\,-\,5\,ng\,ml^{-1}$) over a time course of 8 days. Medium
containing fresh TGF--$\beta$ was replenished every 2 days. Cells
were harvested by trypsinization every $24\,h$, and subjected to
total cell number counting using a Coulter counter. In Figure \ref{Fig4} \textbf{B} each data point represents the mean of 3 wells. Standard deviations were less than 5\% in all cases and were not included in the figure.
For random
motility assay of individual cells, cells were seeded on 6--well
plates at $3\cdot10^4$/well one day prior to image recording. For
motility of cell clusters, cells were seeded at $10^4$/well two days
prior to image recording. Culture media was replaced with fresh
serum--free or full L--15 medium and different concentrations of
TGF--$\beta$  was added $1\, h$ before the image recording. Cells
were imaged on a Nikon TE 2000--E microscope and captured using a
Hamamatsu Orca ER camera and Metamorph software. Images were taken
every 5 minutes for $270$ minutes for random motility assays and
every 30 minutes for 14 hours for the cell cluster experiments.  Image analysis was performed using Metamorph software (MDS, Inc.,
Toronto, Canada). 

\section{Results} 
\subsection{Random motility of individual cells}

\begin{figure}[th]
\begin{center}
\textbf{A}
\includegraphics[width=55mm]{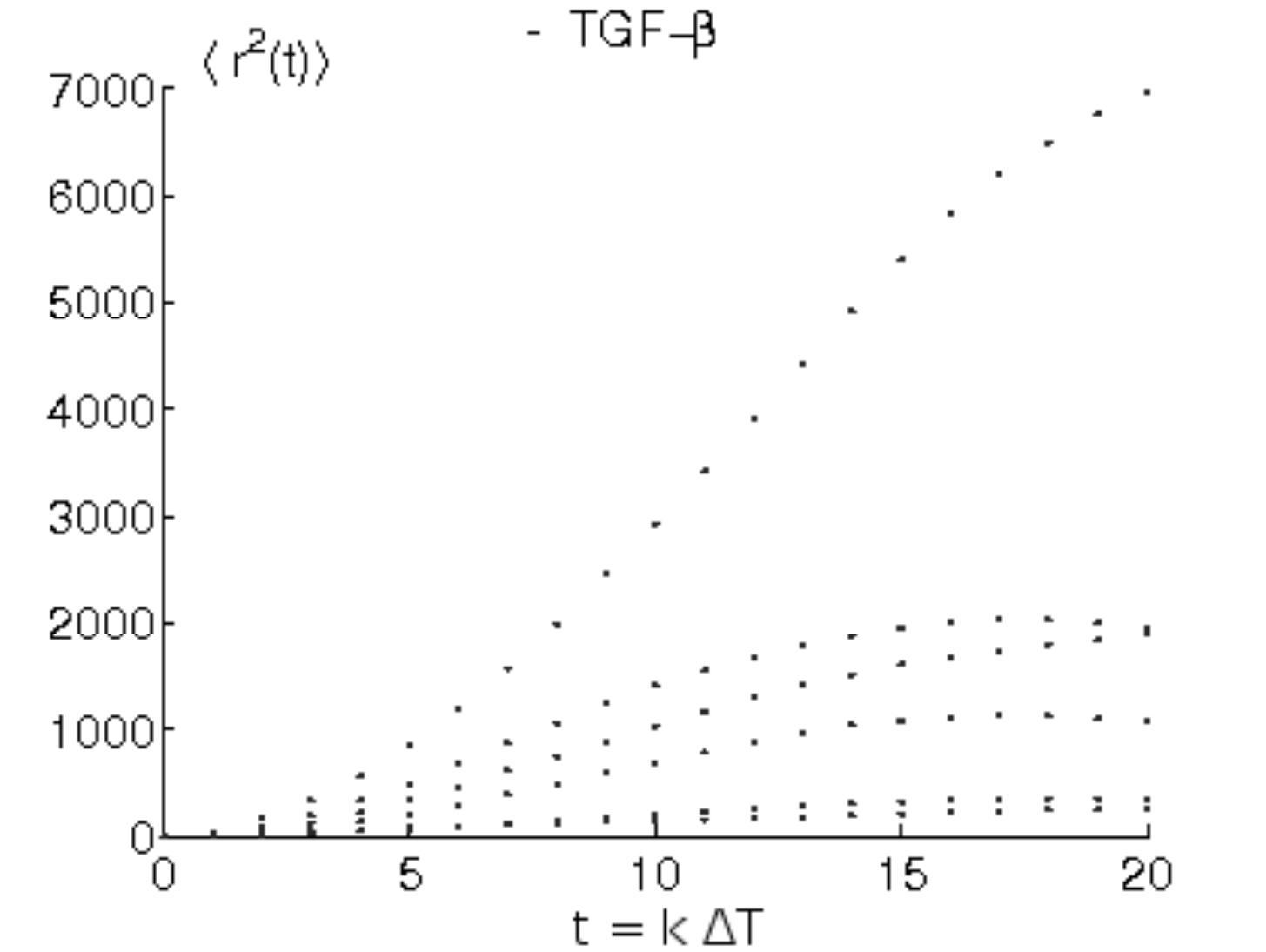}
\textbf{B}
\includegraphics[width=55mm]{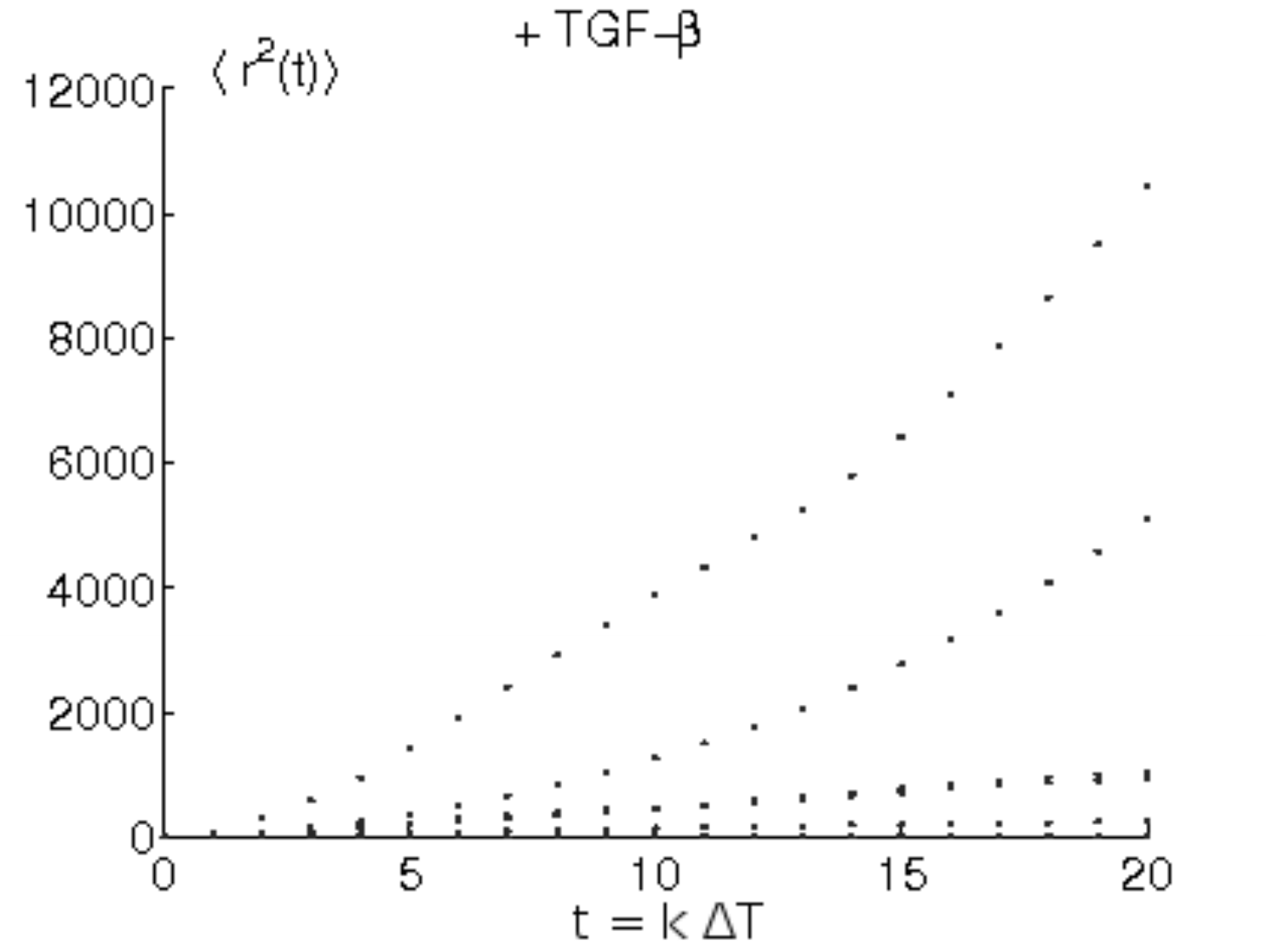}
\caption{\textbf{A} Selected $r^2$ vs.~$k$ curves for untreated
cells.  Position measurements of cells were taken and the mean--squared displacement was calculated
according to equation \eqref{msd}. Out of 20 available curves six showed a roughly linear
dependence of $r^2(k)$. \textbf{B} Selected $r^2$ vs.~$k$ curves for
cells treated with TGF--$\beta$. The time unit is 5
$min$ for both panels, while the mean--squared displacement is in $\mu m^2$.}\label{Fig1}
\end{center}
\end{figure}

We have taken position measurements of individual cells and have
obtained time series data $(x_i,y_i)_{i=1}^N$, with $\Delta
T=5\,min$. The mean--squared displacement (MSD) is calculated
according to
\begin{equation}\label{msd}
   r^2(k \Delta T) = \frac{1}{N-k}\sum_{i=1}^{N-k} \left((x_{i+k}-x_i)^2+(y_{i+k}-y_i)^2\right),
\end{equation}
where $k=1,2,\dots,k_{max}$. The maximal multiple $k_{max}$ for which $r^2$ is calculated has to be chosen carefully. Obviously, the bigger $k$, the fewer such displacements are contained in the time series of length $N$ thus the larger the uncertainty in the estimate $r^2(k)$. The paper \cite{Qian} gives estimates of the standard deviation of $r^2(k)$ in terms of $N$ and $k_{max}$. To obtain the random motility of a single cell we fit the estimate \eqref{msd} with a linear function (see \cite{Berg} for a discussion of this relation)
\begin{equation}\label{lin_growth}
   r^2(k\Delta T) = 4D_mk\Delta T.
\end{equation}
Unfortunately, the analysis is made difficult by cells that seem to remain immobile or whose $r^2(k)$--curves do not look linear. For the plots in  Figure \ref{Fig1} we have selected only curves that could be fit reasonably by a straight line. Other curves show cells
moving initially and then coming to a standstill. Moreover, we cannot see a clear increase of $D_m$ as TGF--$\beta$ is added. For the
average random motility of individual isolated mobile cells we obtain, both for cells
treated with $0$ and with $5\,ng\,ml^{-1}$ TGF--$\beta$,
\begin{equation*}
 D_m = 120\, \mu m^2 h^{-1}.
\end{equation*}
This value can be taken as an upper bound which is realised only if all cells are mobile.

We now look at the percentage of mobile cells. We define a cell to be mobile if it moves outside of a $100\,\mu m \times 100\,\mu m$ square centred at the cell's original position. It becomes clear (Table \ref{Tab1}) that the percentage of mobile cells in our experiment increases as TGF--$\beta$ is introduced.

It seems to be a frequently observed phenomenon that a certain subpopulation of cells remains immobile. Selmeczi \textit{et al.}
\cite{Selmeczi} observed different types of keratinocytes and fibroblasts and selected trajectories only for cells that were
moving at all. This selection procedure is clearly admissible for single particle tracking experiments, but is not applicable to a population with a high fraction of non--mobile cells. We want to emphasize that the cells are not subject to a directional gradient of TGF--$\beta$, rather a uniform concentration applied from all directions, therefore they are able to move randomly in any direction, as opposed to being stimulated to move in one particular direction. Although it is unknown why cells suddenly stop moving, it is possible that in the absence of a strong directional signal there is ``noise'' in the motility signaling that leads to immotility in randomly moving cells.

\subsection{Random motility of cell clusters} 
\begin{figure}[th]
\begin{center}
\includegraphics[width=120mm]{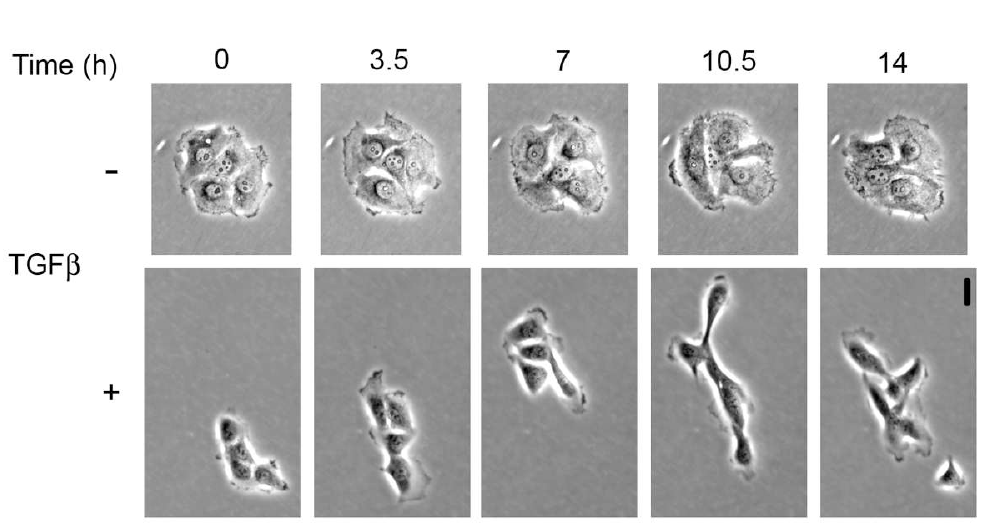}
\caption{Random motility of cell clusters in the absence and presence
 of TGF--$\beta$. The movies are represented by 5
different frames at the indicated times. The bar equals $10\,\mu
m$.}\label{Fig2}
\end{center}
\end{figure}

\begin{figure}[th]
\begin{center}
\textbf{A}
\includegraphics[width=55mm]{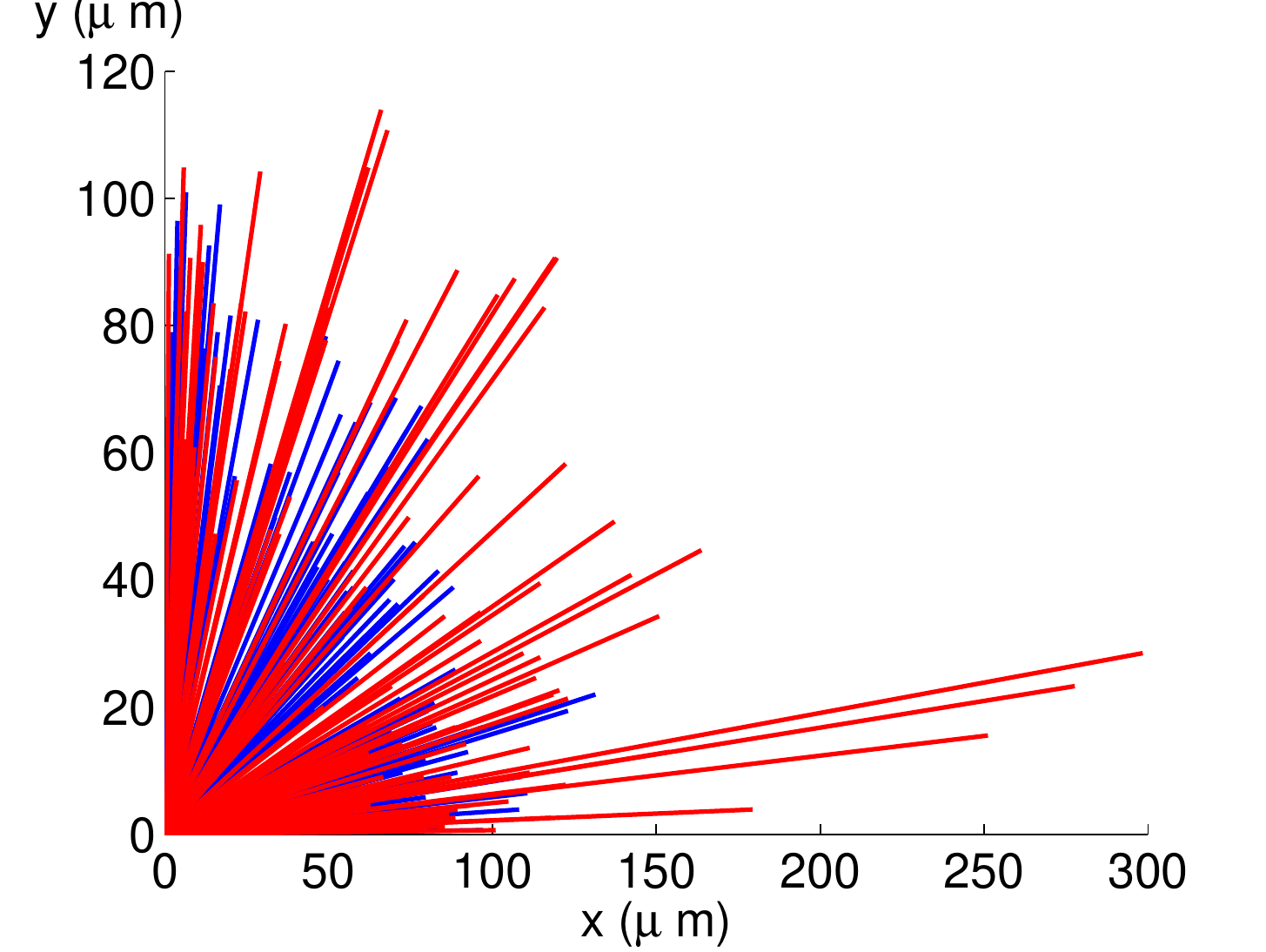}
\textbf{B}
\includegraphics[width=55mm]{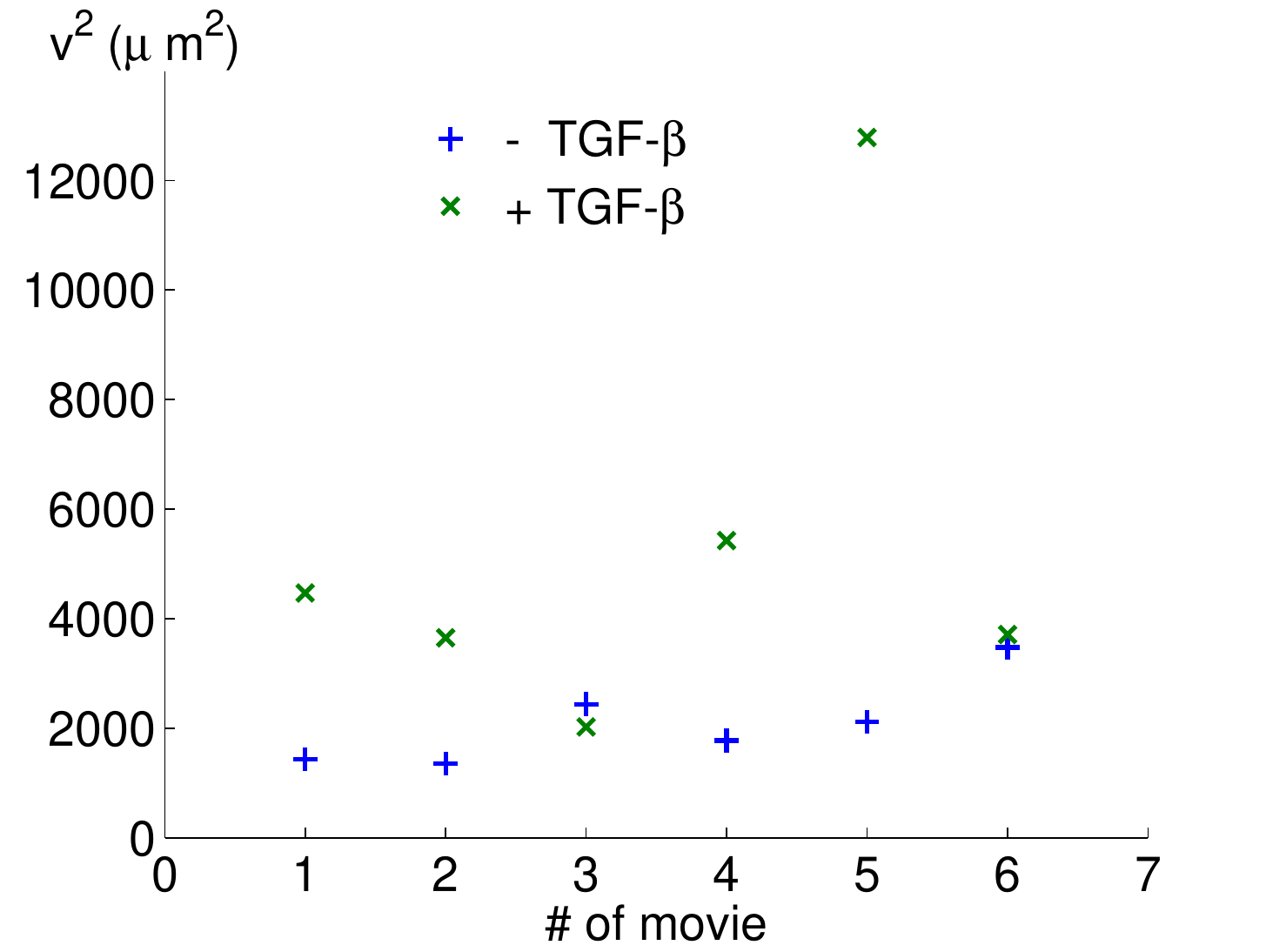}
\caption{\textbf{A} Coverage radii for cell clusters in absence
(blue) and presence (red) of TGF--$\beta$. The fact
that all vectors point into the first quadrant is due to a minor
technicality. \textbf{B} The variation of the squared radii is
computed according to equation \eqref{var} for six different movies. The control cases are
shown in blue, the treatment cases in red. These figures are
representative for three independent realisations of the
experiment.}\label{Fig3}
\end{center}
\end{figure}

We next investigated the motility of cell clusters ($\approx$ 10 cells) in the absence or presence of TGF--$\beta$. As shown in Figure \ref{Fig2}, top row, untreated cell clusters are relatively immobile and seem to maintain a certain shape, although the cells can still rotate within the clusters. In contrast, clusters treated with TGF--$\beta$ are significantly more active in altering their shape and migrating towards unoccupied space (Figure \ref{Fig2}, bottom row). We analysed the movie data in the following way. On the initial frame of the movie the centre of the cluster is identified as the centre of the smallest rectangle containing the cluster. Then, on each subsequent frame of the movie, a straight line is drawn to the point on the boundary of the cluster the farthest from the initial centre. The time gap between two frames is $\Delta T=1\,h$. We obtain a star of radii as shown in Figure \ref{Fig3} \textbf{A}. For a time series of $N$ radii $(r_i)_{i=1}^N$ we calculate the variation of the squares
\begin{equation}\label{var}
   v^2 = \frac{1}{N-1}\sum_{i=1}^{N-1} |r_{i+1}^2-r_i^2|.
\end{equation}
For a total of $36$ movies, $18$ control cases and $18$ treatment cases ($5$ or $10\,ng\,ml^{-1}$)  respectively, a difference can be seen in the value of the variation $v^2$ (Figure \ref{Fig3} \textbf{B}) with TGF--$\beta$ treated cells showing a higher value than untreated cells. Movies were obtained both under
serum--free conditions and with serum. The clusters treated with
TGF--$\beta$ are more active by a factor of two.

We combined the observations of individual mobile cells and clusters of cells to obtain values of $D_m = 5\, \mu m^2 h^{-1}$ (TGF--$\beta$ concentration $0$) and $D_m = 10\,\mu m^2 h^{-1}$ (TGF--$\beta$ concentration $5\,ng\,ml^{-1}$). These values are scaled downwards from the upper bound obtained from the single cell motility experiments  in view of the fact that only a fraction of cells are fully mobile. The ratio of the $D_m$'s for treated and untreated case is suggested by our experimental observations.

\subsection{Growth assays and their numerical simulation}
The mathematical model from equation \eqref{KPP2} was programmed
using \textsc{matlab} (version 7.1, The MathWorks, Inc., Natick,
MA). A standard Crank--Nicolson scheme is used for the diffusion
step \cite[Section 17.3]{Press}, combined with an Euler forward
reaction step. The codes will be available upon request from the
corresponding author.

We fix the following length and time scales for our simulations. The square domain has side length $L=300\,\mu m$ and the time unit is $T = 1\,h$. We assume that a single cell occupies an area of $\beta=100\,\mu m^2$, thus the field has a carrying capacity of $900$ cells. We impose periodic boundary conditions to account for the fact that our numerical domain is a small section of a larger domain, say a Petri dish. Thus, a cell mass that leaves the region will be balanced by a cell coming in at the opposite side. It should also be pointed out that the carrying capacity is the same in control and treatment cases.

\begin{figure}[th]
\begin{center}
\textbf{A}
\includegraphics[height=80mm,width=110mm]{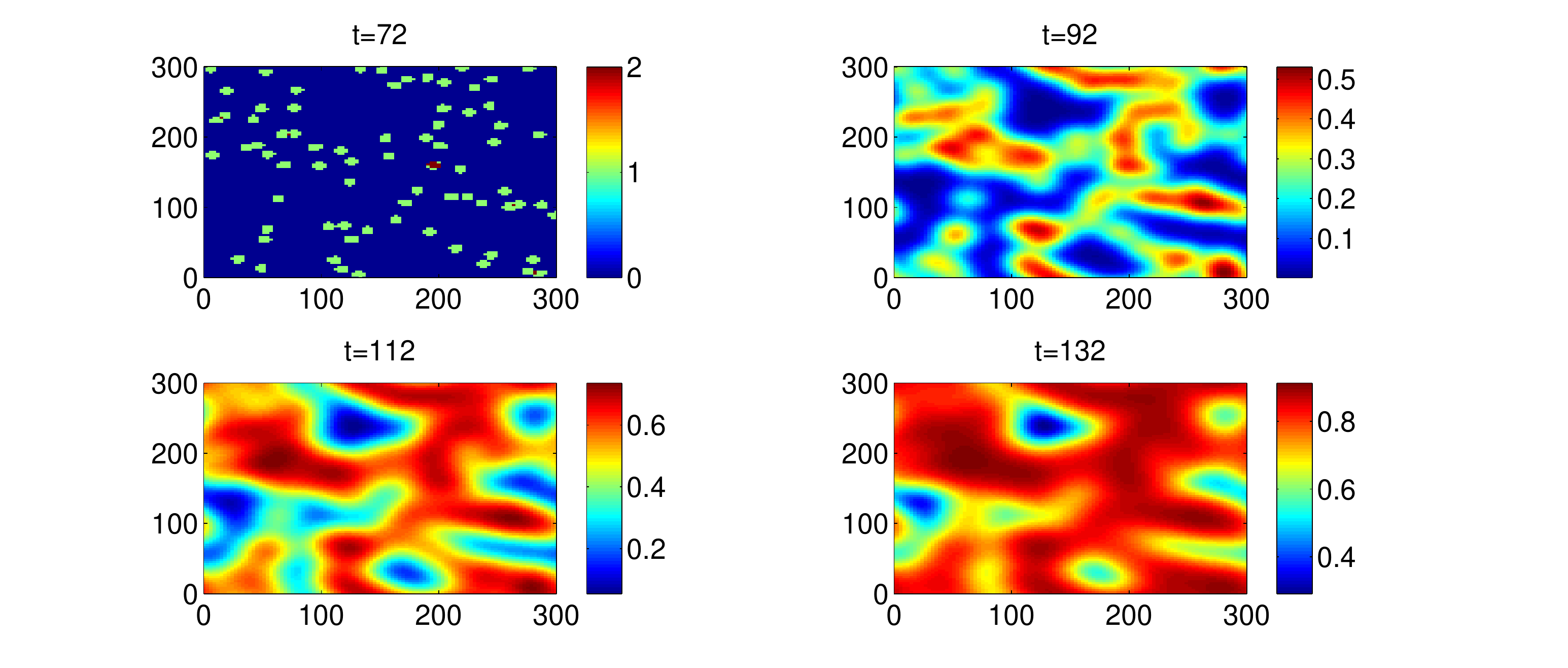}

\textbf{B}
\includegraphics[width=60mm]{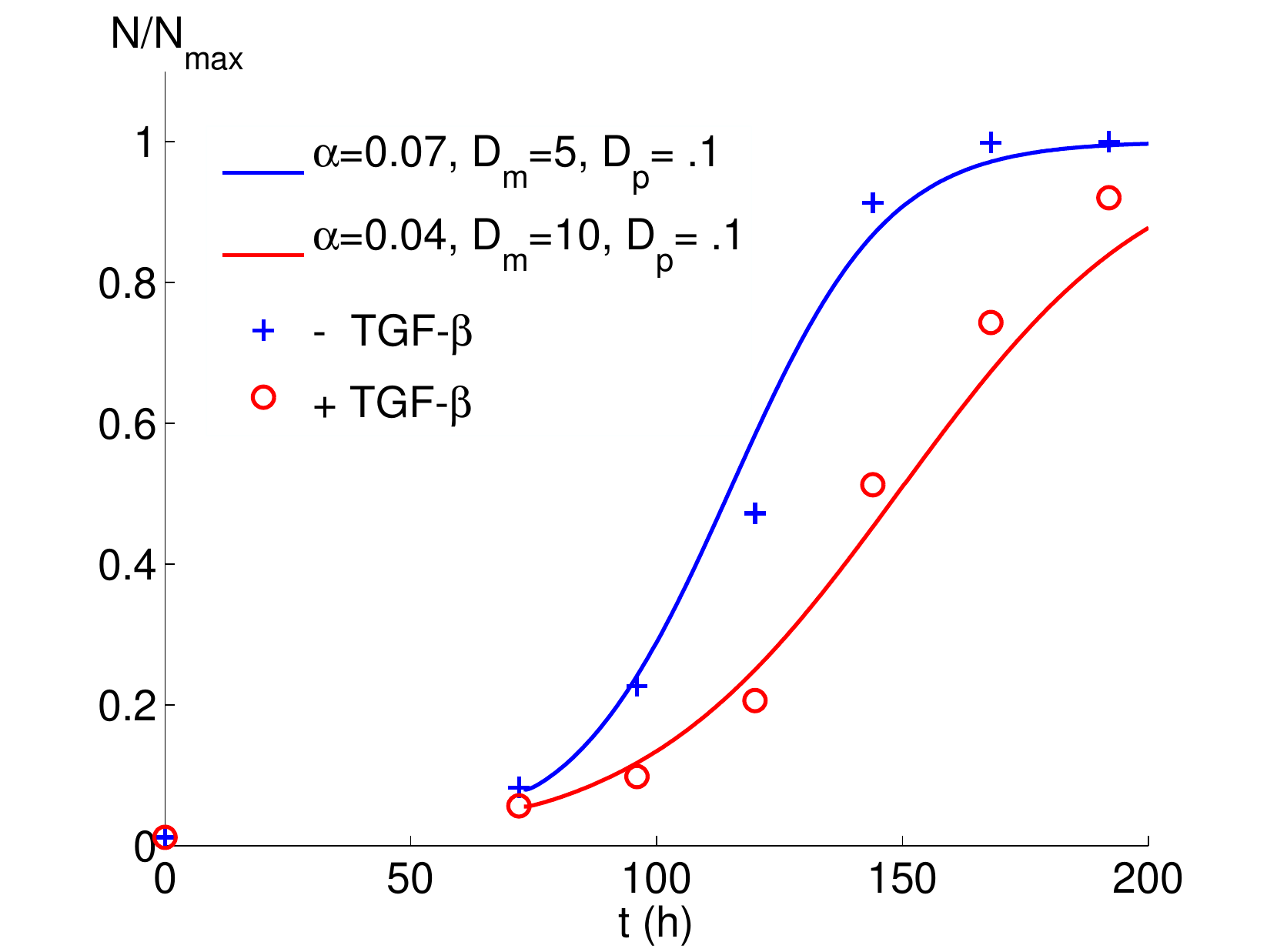}
\caption{\textbf{A} Simulation of untreated cells growing according to the modified
Fisher--Kolmogorov equation \eqref{KPP2} in a  square subfield
of the culture dish with side $300\,\mu m$. Periodic boundary conditions are assumed to account for a balance of cells entering and exiting the subfield. Initially, 75 cells were seeded randomly across
the domain. Snapshots are taken at $t=72\,h$ (initial datum, upper
left) and $t=92,\, 112\,h$ and $t=132\,h$. The parameters are
$D_m=5\,\mu m^2h^{-1}$, $D_p=0.1\,\mu m^2 h^{-1}$ and $\alpha =
0.07\,h^{-1}$. \textbf{B} Normalised growth curves (solid curves, simulation) and experimental measurements
(discrete symbols) over a time course of 5 days. Shown are the
untreated cells (control, $+$) and cells treated with TGF--$\beta$ ($\circ$). }\label{Fig4}
\end{center}
\end{figure}

Numerical simulations of cells growing in a field are shown in Figure \ref{Fig4} \textbf{A}. The initial datum resembles the random placement of cells on the Petri dish. Using our modified Fisher--Kolmogorov equation \eqref{KPP2} we can simulate the experimentally determined growth data (Figure \ref{Fig4} \textbf{B}, discrete symbol represent experimental data, continuous curves represent simulations). At first, the experimental growth data were normalised with respect to a maximum capacity of $1.1\cdot 10^5$ cells.  The parameter $\alpha $ was obtained from fitting the experimental data (see Fig. \ref{Fig4} \textbf{B}), while $D_m$ and $D_p$ were chosen as stated above. The numerical simulation starts only $72\,h$ into the experiment. In our previous experiments with these cells \cite{our_TBMM}, we have noticed that after the cells suspended in medium are seeded onto the plates, they require about 48 hours to adhere to and spread on the plates and adjust to their new environment, before they start a typical growth regime. The numerical parameter values are collected in table \ref{Tab2}. To support our choice of $\alpha$ we note that if the cells were able to grow exponentially according to the law $u'=\alpha u$, then the corresponding doubling times $T_2 = \ln 2/\alpha$
would be $T_2^-=10\,h$ for untreated cells ($\alpha=0.07\,h^{-1}$) and $T_2^+=17\, h$ for cells treated with TGF--$\beta$  ($\alpha=0.04\,h^{-1}$). These doubling times are feasible for MCF10A/HER2 cells in the early exponential growth phase. 

\section{Discussion and Conclusions}
Our experiments demonstrate that TGF--$\beta$ increases the percentage of mobile cells in a cell population in a dose--dependent manner, rather than increases the mean square migration displacement of individual cells.  As we demonstrated, random migration of cell clusters containing $\approx 10$ cells is a feasible approach to parametrise the unbiased random cell migration in a large population of cells. The cluster motility assay indicates that clusters are more mobile and/or less cohesive if TGF--$\beta$ is present. In fact, biological evidence for both mechanisms has been identified. TGF--$\beta$ induces a PI3K--mediated activation of Rac1/Pak1 pathway, which leads to increased cytoskeleton reorganisation, turnover of focal adhesion, and eventually cell migration \cite{27}. Furthermore, TGF--$\beta$ downregulates cell--cell adhesion by decreasing the adhesion protein E--cadherin \cite{28}. The cluster motility assay reflects an overall effect of the multifunctional molecule TGF--$\beta$ on motility. In metastatic cancer cells, TGF--$\beta$ stimulates epithelial--mesenchymal transition (EMT) which is necessary for the acquisition of invasive and metastatic phenotype. It has been shown that in cell monolayer wounded with a pipette tip, the presence of TGF--$\beta$ in the medium markedly induces the closure of wounded area, which is another indicator of TGF--$\beta$--induced cell motility. However, this assay is not used in our parametrisation as the cell migration is not random but targeted to the wounded area.

Our model has a relatively small number of parameters: $D_p$, $\beta$, $D_m$, and $\alpha$. Our procedure for determining these parameters is as follows: (1) $D_p$ and $\beta$ are determined by the average doubling time and average area of cells in the culture, respectively. (2) $D_m$ is determined by adjusting the motility measurements of isolated mobile cells and clusters of mobile cells for their fraction of the total population as the culture attains confluence. From our experiments the upper estimate on $D_m$ was $120 \,\mu m^2 h^{-1}$, which we adjusted downward by $\approx$ 1 order of magnitude to compensate for the fraction of mobile cells in the total population. Table \ref{Tab1} provides the fractions of mobile cells for varied concentrations of TGF--$\beta$. (3) $\alpha$ is determined by fitting model simulations, with the parameters $D_p$, $\beta$, and $D_m$ as above, to the total population data (see Figure \ref{Fig4} \textbf{B}).

It is possible that different values for $D_m$ and $\alpha$ give rise to very similar total population growth curves. We determine ratios of $D_m$ and $\alpha$ for treatment and control cases that simulate our experimental growth data (Figure \ref{Fig4} \textbf{B}). Our determination of  $D_m$  (corresponding to motility) and of $\alpha$ (corresponding to proliferation) reveals that TGF--$\beta$ roughly doubles the value of $D_m$ and halves the value of $\alpha$. The total population growth data of the cell cultures alone (without any further information) are insufficient to quantify these two separate effects of  TGF--$\beta$. We have used in addition the motility experiment data to distinguish the two effects. 

We have developed a general spatial model of proliferating cell cultures \textit{in vitro}, which allows quantification of the properties of proliferative capacity, cell mobility, and clustering as the population attains confluence. The novelty of our interpretation of the Fisher--Kolmogorov equation is that cells begin as isolated geometric regions corresponding to seeding. As they proliferate and migrate these regions expand as controlled by the diffusion parameters and proliferation parameter. When the density is identically $\beta^{-1}$ (whose value is readily correlated to the average area occupied by an individual cell), the region occupied is completely populated. When the density is $<\beta^{-1}$ in a region, then the integral of the density over this region is the expected value of the number of cells in the region. Our focus here is on quantification of the effects of TGF--$\beta$ on motility and proliferation for \textit {in vitro} cancer cell cultures, but our model allows investigation of similar phenomena for many cell types, both prokaryotic and eukaryotic. Extensions and modifications of our model could incorporate different cell proliferation dynamics, multiple cell types and cell interactions (for example endothelial cells, hematopoietic cells, and immune cells), in both \textit{in vitro} and \textit{in vivo} settings. In the future we plan to add more components that naturally exist in the tumour microenvironment to the model, including stromal fibroblasts, immune cells and vascular endothelial cells, which are all affected by TGF--$\beta$. In addition, many tumours carry key mutations on oncogenes or tumour suppressor genes, which may affect their responses to TGF--$\beta$. For example, cancer cells can become unresponsive to the anti--proliferative function of TGF--$\beta$, while remaining sensitive to TGF--$\beta$--induced motility. Inactivating mutations of TGF--$\beta$ receptors are frequently found in pancreatic, biliary and colon cancers \cite{Goggins,Markowitz,16}. Usually, only a subset of cancer cells in a tumour harbour these mutations and these cells live together and communicate with other wild--type epithelial cells, fibroblasts and endothelial cells. It will be important to further investigate with the proper assistance from a mathematical model how these cancers are affected by TGF--$\beta$.
The dual function of TGF--$\beta$ as both a tumour suppressor and promoting factor is reflected by its individual effects on inhibiting cell proliferation and stimulating cell motility. Primary tumour growth by active cell proliferation  results in the increase of tumour size. However, during cancer progression, dissemination of tumour cells from primary site into circulation and the subsequent invasion of other organs, in which cell motility plays a critical role, usually determine the grade of malignancy and the outcome of cancer treatment. Therefore, despite the fact that TGF--$\beta$ has been considered a promising therapeutic target to treat certain types of cancer, the uncertainty of the overall effects of TGF--$\beta$ intervention due to the potential risk of derepressed tumour cell proliferation has been an unsolved critical issue. This requires further in--depth understanding of the different aspects of TGF--$\beta$ action through highly quantitative approaches such as mathematical modeling. Our model described here is the first step of encapsulating the multiple functions of a signalling molecule and potential therapeutic target such as TGF--$\beta$ into a computational model. By introducing the modeling method to the cancer biology of TGF--$\beta$, questions that have been raised for years will have the chance to be answered from a novel angle. These interesting topics include: (1) When is the critical time of TGF--$\beta$ being switched from a tumour suppressor to a promoter? (2)  When is the most efficient time to apply anti--TGF--$\beta$ therapy during cancer progression? (3) Will therapeutic inhibition of TGF--$\beta$ at pre--cancer lesions or at early stages of the disease trigger primary tumour growth as a result of derepression of cell proliferation? Besides the complexity of TGF--$\beta$ signalling, cancer cells can respond differently to TGF--$\beta$. For example, several oncogenes including HER2 are known to synergize with TGF--$\beta$ to function on cell motility and survival. Some cancers do not respond to the anti--proliferative effect of TGF--$\beta$ but still respond to its stimulatory effect on motility. The mathematical model will therefore provide a unique way to simulate how TGF--$\beta$ functions differently in cancer cell populations with different characteristics, and the effect of TGF--$\beta$ therapeutic inhibition on a tumour consisting of heterogeneous cell populations.

\section{Acknowledgements} 
This work was supported by the NCI Integrative Cancer Biology Program (U54 CA113007), K22 CA109590 (AMW), K99 CA125892 (SEW), R01
CA80195 (CLA), R01 CA62212 (CLA), Breast Cancer Specialised Program of Research Excellence (SPORE) P50 CA98131, and the Vanderbilt
Integrative Cancer Biology Center (VICBC). The authors thank Dr.~Vito Quaranta for support.

\bibliography{TGFbeta_Emily}

\begin{table}[th]
\begin{center}
\caption{The percentage of mobile cells in the single cell experiments as function of TGF--$\beta$ concentration}\begin{tabular}{c|c}
$c\:(ng\,ml^{-1})$ &  motility (\%) \\
\hline
$0$   & $33$  \\
$0.5$ & $45$ \\
$1$   & $49$ \\
$2$   & $51$ \\
$5$   & $56$
\end{tabular}\label{Tab1}
\caption{The parameter values used in the simulations shown in figure \ref{Fig4} \textbf{B}}
\begin{tabular}{c|c|c|c}
TGF--$\beta$ & $D_m\:(\mu m^2h^{-1})$ & $D_p\:(\mu m^2h^{-1})$ & $\alpha\:(h^{-1})$  \\
\hline
$-$   & $5$  & $0.1$ & $0.07$  \\
$+$   & $10$ & $0.1$ & $0.04$
\end{tabular}\label{Tab2}
\end{center}
\end{table}


\end{document}